\documentclass{iau}

\usepackage{amsmath}
\usepackage{graphicx}
\usepackage{multirow}

\newcommand{\kep}{\emph{Kepler}}

\begin{document}

\lefttitle{Cambridge Author}
\righttitle{Proceedings of the International Astronomical Union: \LaTeX\ Guidelines for~authors}

\jnlPage{1}{7}
\jnlDoiYr{2021}
\doival{10.1017/xxxxx}

\aopheadtitle{Proceedings IAU Symposium}
\editors{A. V. Getling \&  L. L. Kitchatinov, eds.}

\title{Convection, rotation, and magnetic activity of solar-like stars from asteroseismology }

\author{Savita Mathur}
\affiliation{Instituto de Astrof\'isica de Canarias (IAC), E-38205 La Laguna, Tenerife, Spain}
\affiliation{Universidad de La Laguna (ULL), Departamento de Astrof\'isica, E-38206 La Laguna, Tenerife, Spain}

\begin{abstract}
During the last decade, our understanding of stellar physics and evolution has undergone a tremendous revolution thanks to asteroseismology. Space missions such as CoRoT, \kep, K2, and TESS have already been observing millions of stars providing high-precision photometric data. 
With these data, it is possible to study the convection of stars through the convective background in the power spectrum density of the light curves.  The properties of the convective background or granulation has been shown to be correlated to the surface gravity of the stars. 
In addition, when we have enough resolution (so long enough observations) and a high signal-to-noise ratio (SNR), the individual modes can be characterized in particular to study the internal rotational splittings and magnetic field of stars. 
Finally, the surface magnetic activity also impacts the amplitude and hence detection of the acoustic modes. This effect can be seen as a double-edged sword. Indeed, modes can be studied to look for magnetic activity changes. However, this also means that for stars too magnetically active, modes can be suppressed, preventing us from detecting them.

In this talk, I will present some highlights on what asteroseismology has allowed us to better understand the convection, rotation, and magnetism of solar-like stars while opening doors to many more questions. 
\end{abstract}

\begin{keywords}
Asteroseismology, Convection, Rotation, Magnetic activity, Solar-type stars
\end{keywords}

\maketitle

\section{Introduction}

While the study of the Sun with both local and global helioseismology has been going on for many decades \citep[e.g.][]{1979Natur.282..591C}, the possibility to study other stars like the Sun has been possible on a very large scale thanks to the ESA and NASA space missions such as CoRoT \citep[Convection, Rotation, and planetary Transits,][]{2006ESASP.624E..34B}, {\it Kepler} \citep{2010Sci...327..977B}, K2 \citep{2014PASP..126..398H}, and TESS \citep[Transiting Exoplanet Survey Satellite,][]{2015JATIS...1a4003R}. Most of these missions focus on the search for exoplanets but the importance of characterizing the star at the center of planetary system has been part of the main objectives of CoRoT and the more recent missions. Indeed the high quality of the photometric data collected by such missions allowed us to do asteroseismic analyses with many new discoveries that are revolutionizing our understanding of stellar evolution structure and dynamics. 

The ESA M3 mission, PLATO \citep[PLAnetary Transits and Oscillations of stars, ][]{2014ExA....38..249R}, which launched is scheduled for the end of 2026 will continue in this direction with the goal of characterizing stars like the Sun fo spectral types F, G, and K on the main sequence and increase by more than a factor of 100 the current {\it Kepler} Legacy catalog \citep[e.g.][]{2011A&A...534A...6C,2011ApJ...733...95M,2012ApJ...748L..10M,2015MNRAS.452.2127S,2017ApJ...835..173S}.

Finally, the ESA Roman Space telescope \citep{2023ApJS..269....5W} with a launch scheduled in May 2027 and  whose main objectives will be to study dark energy and dark matter, as well as image exoplanets will provide data that will be well suited for asteroseismic analyses \citep{2023arXiv230703237H}.

Up to now, \kep\, is the mission that provided the largest sample of solar-like main-sequence and subgiant stars with detected solar-like oscillations \citep{2014ApJS..210....1C,2016MNRAS.456.2183D,2017ApJ...835..172L,2022A&A...657A..31M} with more than 620 stars. The K2 mission that followed the nominal \kep\, mission after the second reaction wheel failed also contributed to increase the sample of solar-like stars with seismic detections \citep[e.g.][]{2016PASP..128l4204L,2021ApJ...922...18O,2023A&A...674A.106G}. Finally, while the TESS mission has been observing millions of stars since its launch in 2018, it's observing feature and an instrumental noise level higher the one of  \kep\, did not lead as many detections of solar-like oscillations in main sequence and subgiant stars \citep[e.g.][]{2019AJ....157..245H}. The new 20-second cadence improved the signal-to-noise ratio \citep{2022AJ....163...79H,2023A&A...669A..67H} and the yield of detections should increase with the new cycles.

The path for asteroseismic studies has been well paved and is very bright for the near future. 

When computing the power spectrum density (PSD) computed from photometric observations contains information on several physical processes in stars: surface rotation, convection, and oscillations \citep{2019LRSP...16....4G}. In the next sections, we will go through the different highlights on convection, rotation and magnetism of stars thanks to asteroseismic analyses. 


\section{Convection}

The continuum in the PSD of a solar-like stars results from the convection (or granulation) occurring in the outer layers of solar-like stars \citep{2019LRSP...16....4G}. That convective background is generally modeled with several components by many of the asteroseismic pipelines. For the Sun, \citet{1985ESASP.235..199H} proposed the following function to model the granulation:

\begin{equation}\label{eq:Ro}
     B = \frac{4 \sigma^2 \tau_{\rm gran}}{1 + (2\pi\nu\tau_{\rm gran})^\alpha},
 \end{equation}

\noindent where $\sigma$ is the characteristic amplitude of the granulation, $\tau_{\rm gran}$ the characteristic time scale of granulation. For the Sun $\alpha$\,=\,2 but \citep{2014A&A...570A..41K} showed that for other stars where we model the background with two Harvey-like functions -- corresponding to different scales of granulations -- a slope $\alpha$\,=\,4 was favored.
By studying the granulation of several hundreds of red giants and main-sequence solar-like stars observed by \kep, a correlation between the surface gravity and $\sigma$ and $\tau$ was found \citep{2011ApJ...741..119M,2014A&A...570A..41K}. The analysis also showed that the granulation timescale increases with the evolution of the star. 

The comparison of the \kep\, observations of granulation with 3D simulations of convection in a box by  \citet{2011ApJ...731...78T}emphasized that the simulations underestimated the timescale by a factor of 2 while they overestimated the granulation power by an order of magnitude. These differences could be explained by the fact that the grid of 3D simulations was sparse and extrapolations to the stellar parameters of the observations were performed and also by the fact that the simulations were done for solar metallicity, which can impact the stellar convection. More studies are being done where metallicity from spectroscopic surveys are used (Perillo et al. in prep.).

The strong correlation between surface gravity and granulation led to the definition of different metrics to infer the $\log g$ of stars. The first one is the Flicker \citep{2013Natur.500..427B} computed on a light curve smoothed over 8 hours. But that metric is limited to a small range of red giant stars and can be biased for stars with oscillations close to the 8-hour time scale. 

\citet{2018A&A...620A..38B} developed another metric computed on the PSD: the FliPer (Flicker in Power), which is the total power in the PSD between a given minimum frequency ($f_{\rm min}$ and the Nyquist frequency to which the photon noise is subtracted. The photon noise for the \kep\, data is computed following the relation from \citet{2010ApJ...713L.120J}. The FliPer computed for the \kep\, targets showed a very good correlation with $\log g$.

In order to avoid any issues from low-frequency peaks related to the surface rotation or spurious spikes, the FliPer is computed with several $f_{\rm min}$ values. These values along with the effective temperature and  \kep\, magnitude were then used as input of a Random Forest algorithm \citep{2001MachL..45....5B}. After training the algorithm, the test set showed that the RF provided $\log g$ values with uncertainties of 0.046\,dex on average, much better that spectroscopic uncertainties and close to the seismic ones. 

This metric is very powerful, especially because the modes do not need to be detected. So it can be applied to stars with short light curves and  with low SNR. It can also be used to classify stars \citep{2019A&A...624A..79B}.

\section{Rotation}

In this section, we will focus on the rotation of stars from the main-sequence to the red giant branch.

\subsection{Surface rotation}

The need to understand the transport of angular momentum in stars is crucial as it impacts the mixing of elements and hence the ages of stars. We will see how asteroseismology can help understand the evolution of surface rotation with age. In 1972, \citeauthor{1972ApJ...171..565S} studied two clusters and the Sun and showed that the surface rotation rate decreased as a square root of the age.  This was later generalized with more clusters and field stars and led to the definition of rotation-age relations, also known as gyrochronology \citep{2007ApJ...669.1167B}. 

With the high-quality of the \kep\, data, surface rotation periods could be measured through the modulation of the light curves due to the presence of spots and active regions \citep[e.g.][]{2013A&A...557L..10N,2013A&A...560A...4R,2014ApJS..211...24M,2014A&A...572A..34G}. In addition, for a subsample of stars, asteroseismic analyses were possible, which provided ages of stars \citep[e.g.][]{2012ApJ...749..152M,2014ApJS..214...27M,2017A&A...601A..67C, 2017ApJ...835..173S}. Seismic ages are obtained by fitting spectroscopic (effective temperature, metallicity) and seismic observables (including global seismic parameters and individual frequencies of the modes) with stellar models. \citet{2014A&A...569A..21L} showed that by including the individual frequencies of the modes, the inference of the masses, radii, and ages of the stars were improved with a better precision. 

By combining surface rotation periods with precise seismic ages, \citet{2016Natur.529..181V} compared the evolution of rotation with age of solar-like stars with models including gyrochronology prescriptions. They found that stars older than the Sun were rotating faster than expected given their ages. This discovery suggested that at a given moment (close to the Rossby number of the Sun) during the main-sequence life of a star, the magnetic braking weakens. This means that gyrochronology is only applicable to stars below that Rossby threshold. It was also noticed that the Sun seems to be closed to that transition phase. The exact origin of that phenomenon is not yet understood but hints on change in the differential rotation and magnetic topology are being investigated \citep[e.g.][]{2020ApJ...900..154M}.

\subsection{Latitudinal differential rotation}

Internal rotation of the Sun and stars can be measured through rotational splittings of the acoustic modes. Indeed, rotation lifts the degeneracy and yields several 2l+1 components of modes with degree l above 1. The extracted splittings also depend on the inclination angle of the stars as they impact which components are visible \citep{2006MNRAS.369.1281B}.

For the \kep\, solar-like targets with the highest SNR, several works looked for signature of latitudinal differential rotation \citep[e.g.][]{2014ApJ...790..121L,2016A&A...586A..79S} by using rotational splittings measurement. They mostly concluded that the current data did not allow to have a firm measurement of differential rotation due to the small number of low-degree modes detected.

We highlight her the work by \citet{2018Sci...361.1231B} who looked at the Clebsch-Gordon $a$ coefficients of the rotation splittings. The third coefficient, $a_3$ contains information on the latitudinal differential rotation. They measured that $a_3$ coefficient for 40 solar-like stars observed by \kep\, and computed the detection probability for stars to have solar or anti-solar rotation. They found 13 candidates with the most significant detection with a solar differential rotation. 

\subsection{Internal rotation}

For more evolved stars, the detection of mixed modes that result from the coupling of acoustic- and gravity-mode cavities allowed the study of the core rotation of subgiants and red giants \citep{2012Natur.481...55B,2014A&A...564A..27D}. The cores of these evolved stars were found to rotate faster than the envelopes on average.   While \citet{2012A&A...548A..10M} suggested that the core of red-giant branch (RGB) stars were spinning down towards the red clump (RC) stars, an extended analysis by \citet{2018A&A...616A..24G} did not confirm such trend. The cores of red-giant stars seem to be constant on the RGB without dependence on mass. The spinning down from RGB to RC can partially explain the expansion of the non-degenerated helium core but not the full slow down.

\section{Stellar magnetism}

Stellar magnetism can also be studied with asteroseismology for both surface and internal magnetism. 

\subsection{Impact of magnetic activity on solar-like oscillations}

For the Sun, it is well known that during its magnetic activity cycle, the acoustic modes are affected in the following way: while magnetic activity increases, the frequencies of the modes increase and their amplitudes decrease \citep[e.g.][]{1985Natur.318..449W,1990Natur.345..322E,2009ApJ...695.1567J}. Similar behavior was first observed in another solar-like stars, HD~49933, observed by the CoRoT mission \citep{2010Sci...329.1032G}. An anti-correlation between the temporal changes of the amplitude and the frequency shifts of the modes was observed. Complementary CaHK observations at the Cerro Tololo observatory done a few years later confirmed that the star was active with an S-index of 0.31. Later with \kep, such phenomenon was detected in more than 50 targets \citep[e.g.][]{2016A&A...589A.118S,2017A&A...598A..77K,2018ApJS..237...17S} where changes were correlated with the photometric activity index, $S_{\rm ph}$ \citep{2014A&A...562A.124M}. These detections can provide important constraints on the mechanisms involved in the dynamo of these stars. 

One example is the \kep\, target KIC~8006161 for which frequency shifts and changes in the amplitudes of the modes were detected. That star was also followed up decades before the \kep\, observations with the Keck providing chromospheric index of magnetic activity, in particular a detection of a magnetic activity cycle of 7.4 yrs. From seismology the stellar parameters of KIC~8006161 are very similar to the Sun. Having a stronger chromospheric emission than the Sun and a shorter cycle period could be explained with the fact that KIC~8006161 is a meta-rich star. \citet{2018ApJ...852...46K} showed how the metallicity impacted the magnetic activity of the star.

\citet{2019ApJ...883...65S} studied more thoroughly the sample of \kep\, targets with significant frequency shifts as they were also modeled with seismology. As expected, magnetic activity decreases with age \citep[e.g.][]{1978ApJ...226..379W}. Magnetic activity was also found to be stronger with higher metallicity. 

Since a higher level of magnetic activity leads to smaller amplitude of the acoustic modes, this also means that stars with a very high magnetic activity will have much smaller amplitudes of the modes \citep{2011ApJ...732L...5C}.  

The rotation study of $\sim$\,1000 \kep\, stars observed in short cadence, where mode detection was expected but was not achieved, yielded a sample of 323 stars with reliable rotation periods \citep{2019FrASS...6...46M}. The results were compared to the ones of $\sim$\,300 stars with solar-like oscillations detected and reliable rotation.  Among the stars without detected oscillations, 32\% have a photometric magnetic activity index, $S_{\rm ph}$, larger than the maximum level of the Sun, explaining the non detection of the modes. However, there remains a large fraction of the sample with magnetic activity similar to the Sun and for which modes were not observed. The reason behind the non-detection is not clear but hints are possibly: metallicity and impact of the inclination angle of the rotation axis. But larger sample with spectroscopic observations is needed to confirm such trends.

\subsection{Internal magnetic field of evolved stars}

Another sample of stars with unusual mode pattern was highlighted by the \kep\, mission: stars with abnormal low-amplitude dipole modes \citep{2012A&A...537A..30M,2014A&A...563A..84G} in reg-giant stars. One of the theories developed suggests that there is a strong magnetic field in the radiative zone of those stars, which interacts with the l=1 modes and scatters them into high-degree modes that are trapped in the radiative zone \citep{2015Sci...350..423F}. The typical critical magnetic field responsible for such phenomenon was estimated to $10^5$\,G. The analysis of a larger statistical sample showed that stars with these {\it depressed} modes had a mass above 1.2\,M$_\odot$, corresponding to stars that developed a convective core \citep{2016Natur.529..364S} agreeing with the aforementioned theory.

The same way rotation lifts the degeneracy of the modes, magnetic field was also shown to have such an impact \citep{DuvDzi1984,GooTho1992}. Recently several teams studied the effect of an axisymmetric field on mixed modes \citep{2021A&A...647A.122M,2021A&A...650A..53B,2021MNRAS.504.3711L}.  They showed that this would lead to an asymmetric triplet and suggested that such asymmetry could be searched for in the \kep\, data. The generalization of the theory to non-axisymmetric magnetic field led to the discovery asymmetries in the splittings of red giants attributed to fields of strength 30 to 100\,kG \citep{2022Natur.610...43L,2023A&A...670L..16D}.
\newline


With the coming launch of the ESA PLATO mission in late 2026, even more breakthroughs should be made, helping us in our comprehension of stellar interior dynamics evolution.

\section{Acknowledgements}
S.M.\ acknowledges support by the Spanish Ministry of Science and Innovation with the Ramon y Cajal fellowship number RYC-2015-17697, the grant number PID2019-107187GB-I00, the grant no. PID2019-107061GB-C66, and through AEI under the Severo Ochoa Centres of Excellence Programme 2020--2023 (CEX2019-000920-S).

\bibliographystyle{iaulike}
\bibliography{/Users/Savita/Documents/BIBLIO_sav}

\end{document}